\documentclass[aps,twocolumn,prl,superscriptaddress]{revtex4-2}

\usepackage{amsmath,amssymb}
\usepackage[T1]{fontenc}
\usepackage{graphicx}
\usepackage{dcolumn}
\usepackage{bm}
\usepackage{siunitx} 
\usepackage{xcolor}
\usepackage{listings}
\begin{document}

\title{Polaritonic Machine Learning for Graph-based Data Analysis}%

\author{Yuan Wang}
\affiliation{School of Mathematical and Physical Sciences, University of Sheffield, Sheffield S10 2TN, United Kingdom}

\author{Stefano Scali}
\affiliation{Department of Physics and Astronomy, University of Exeter, Stocker Road, Exeter EX4 4QL, United Kingdom}

\author{Oleksandr Kyriienko}
\affiliation{School of Mathematical and Physical Sciences, University of Sheffield, Sheffield S10 2TN, United Kingdom}

\date{\today}

\begin{abstract}
Photonic and polaritonic systems offer a fast and efficient platform for accelerating machine learning (ML) through physics-based computing. To gain a computational advantage, however, polaritonic systems must: (1) exploit features that specifically favor nonlinear optical processing; (2) address problems that are computationally hard and depend on these features; (3) integrate photonic processing within broader ML pipelines. In this letter, we propose a polaritonic machine learning approach for solving graph-based data problems. We demonstrate how lattices of condensates can efficiently embed relational and topological information from point cloud datasets. This information is then incorporated into a pattern recognition workflow based on convolutional neural networks (CNNs), leading to significantly improved learning performance compared to physics-agnostic methods. Our extensive benchmarking shows that photonic machine learning achieves over 90\% accuracy for Betti number classification and clique detection tasks - a substantial improvement over the 35\% accuracy of bare CNNs. Our study introduces a distinct way of using photonic systems as fast tools for feature engineering, while building on top of high-performing digital machine learning.
\end{abstract}

\maketitle


\textit{Introduction.---}Machine learning is a powerful tool for data processing tasks \cite{LeCun2015}, ranging from natural language processing \cite{vaswani2017attention,brown2020language} to scientific discovery \cite{jumper2021highly,Rudy2017,wang2023scientific}. One part of ML success comes from the development of models that can capture relevant features for different data types \cite{wang2023scientific,tabakhi2023multimodal}. This corresponds to extracting patterns from images with convolutional neural networks (CNNs) \cite{krizhevsky2012imagenet,oshea2015cnns,Strofer2019cnnsforvortex}, link predictions for molecular graphs with geometric machine learning \cite{bronstein2021geometric,zhang2018link}, and tokenized amino acid sequences processed by transformers \cite{jumper2021highly}. Another part of success comes from the growth of computing power (e.g. GPUs---graphical processing units), which allows running deep ML architectures with billions of weights \cite{sevilla2022compute}. At the same time, solutions based on conventional electronics scale extensively, leading to rapidly increasing energy consumption \cite{Nature2025_data_centers}. Proposing nature-inspired and physics-based solutions is considered a way forward for building ML pipelines in the future \cite{muir2025road_neuromorphic,brunner2025_roadmap_neuromorphic_photonics,finocchio2024_roadmap_nanotech,sheik2023_hardware_is_software,Serra-Garcia2019}.

Physical systems based on optics and photonics emerged recently as promising platforms for ML, thanks to their favorable features that include large bandwidth and low-loss operation \cite{mcmahon2023rev}. Based on propagating light, these systems can address various ML tasks bypassing conventional electronics \cite{li2024onn_review, shastri2020neuromorphic_review}. This includes performing classification with diffraction neural networks \cite{lin2018d2nn}, gaining a huge increase in convolutional capacities with frequency combs \cite{feldmann2020onn_nn}, and in-situ backpropagation \cite{wright2022deep_physical_nn,pai2023photonic_backprop}. Optical ML-based sensing capabilities were demonstrated in Ref.~\cite{wang2023image_sensing_onn}, achieving high accuracy and sub‑photon energy per operation \cite{wright2022optonn}. Furthermore, optical neural networks were used for generative modeling \cite{choi2024photonic_probabilistic,bruckerhoff2024chaotic_light,zhan2024photonic_diffractive_generators}. As ML models crucially require nonlinearity, linear optical neural networks often incorporate it by analog-to-digital conversion or intensity saturation \cite{wang2023image_sensing_onn}, showing significant improvement over bare linear performance. Hence, designing physically nonlinear devices for processing shall largely increase ML capabilities.

Nonlinear optical effects depend on coupling between light and matter, and are greatly enhanced when cavity-confined photons hybridize with excitons \cite{basov2025_polaritonic_quantum_matter,Deng2010rev,carusotto2013quantum}. The resulting quasiparticles---exciton-polaritons---exhibit strong nonlinearity per unit volume \cite{DEVEAUD2016,Estrecho2019}, emerging from exciton-exciton scattering and nonlinear phase space filling \cite{Tassone1999,Shahnazaryan2017,Yagafarov2020,KWSong2024,genco2024femto,richard2025}. The nonlinear response can be enhanced in the presence of dipolar interactions \cite{Togan2018,rosenberg2018,datta2022highly_nonlinear_dipolar_polaritons,Louca2023,Christensen2024}, Rydberg states \cite{gu2021enhanced_nonlinear_rydberg_polaritons,Makhonin2024,Shahnazaryan2016}, doping-related effects \cite{emmanuele2020_trion_polaritons,Tan2020prx,Kyriienko2020prl,Wei2023}, moiré patterns for bilayers of 2D materials \cite{Zhang2021,Luo2023,CamachoCooper2022,KWSong2024,herrera2025moirereview,gr72-szwg}, and biexcitonic resonances \cite{Takemura2014,Navadeh2019}. A distinct feature of polaritonic systems is their ability to form condensates as macroscopically coherent states \cite{Kasprzak2006,Schneider2013,Kavokin2022,Han2025}, leading to many-body excitations in the form of quantized vortices \cite{Lagoudakis2008,Amo2009,Sanvitto2010,Krizhanovskii2010,Roumpos2011,Ma2020,Gnusov2023,Alyatkin2024}, phase defects and solitons \cite{Sich2012,Walker2015,Liew2015,Maitre2020}. These features of fluids of light \cite{carusotto2013quantum} can also be paired with rich physics of polaritonic lattices created via patterning \cite{Solnyshkov2021,Kim2013,Jacqmin2014,Whittaker2018,Kyriienko2019,Dusel2020,Betzold2024,Kedziora2024} and optical pumping \cite{Cristofolini2013,Berloff2017,Alyatkin2021,Alyatkin2024b,Zaremba2025}. These capabilities suggest promise for polaritonic ML applications.
\begin{figure*}[t]
\begin{center}
\includegraphics[scale=1.0]{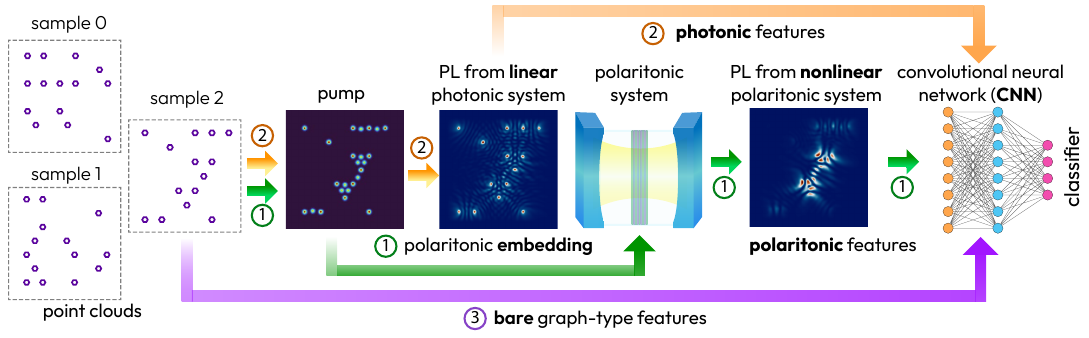}
\end{center}
\caption{Workflow for point cloud classification based on convolutional neural networks (CNNs). Datasets are processed in three distinct ways, using bare graph-type images, photoluminescence (PL) of a linear photonic system, and PL of a nonlinear polaritonic system. Green arrow \raisebox{.5pt}{\textcircled{\raisebox{-.9pt} {1}}}: point clouds are incorporated into a nonresonant pump profile featuring both source and barrier spots, triggering polariton condensation with nonlinear features. Corresponding PL is used in the CNN-based classification. Orange arrow \raisebox{.5pt}{\textcircled{\raisebox{-.9pt} {2}}}: data embedded into a potential profile of linear photonic systems, with PL (light intensity) used as an input for CNNs. Purple arrow \raisebox{.5pt}{\textcircled{\raisebox{-.9pt} {3}}}: bare point clouds are directly used as input for the CNN workflow.}
\label{fig_1}
\end{figure*}

To date, polaritonic approaches to machine learning concern models designed to mimic neural network-type structures, Ising-type optimizers, and reservoir computing (RC) devices \cite{Opala2023rev}. Proposals include polaritonic RC for speech recognintion \cite{Opala2019}, polariton neural networks for MNIST image classification \cite{Ballarini2020} with RC-type offline learning, binarized and time-delay polariton networks for learning optical logical gates \cite{Mirek2021Nano,Mirek2022} and image recognition \cite{Sedov2025}, and recently a full polaritonic four-neuron network for pattern recognition \cite{opala2024rtPNN}. Energy efficiency of polaritonic ML was quantified in Ref.~\cite{Matuszewski2021}. In a parallel track, polaritonic lattices were suggested as effective quantum reservoirs for quantum state discrimination \cite{Ghosh2019,Ghosh2021,Ghosh2021b,Krisnanda2023}---something that becomes within reach for quantum polaritonics \cite{Kuriakose2022,Delteil2019,Munoz-Matutano2019}. The proposed approaches show promise, yet still operate at a small scale as compared to current deep learning.

In this Letter, we propose a polariton-based machine learning evolved in a distinct direction. First, instead of trying to reproduce deep learning with nonlinear photonics, we use polaritonic systems for feature engineeting, while offload post-processing to known classical ML architectures (Fig.~\ref{fig_1}). This is similar to positional embeddings, proven to be state-of-the-art \cite{vaswani2017attention,ying2021transformers,ma2023grit}.  Second, we choose a task that naturally fits polaritonic lattices yet remained unexplored, corresponding to the analysis of graph-type data and point clouds. Here, polaritonic systems can act as physical accelerators used for highlighting certain features that originate from topology in system's response as well as the topology of data \cite{wasserman2016tda}. We show that by introducing nonlinear features from polariton condensation, CNN-based classification can significantly improve on purely graph-based analysis. 


\textit{Model: point cloud and graph analysis.---}We start by describing the data preparation and specifying the learning tasks. Consider a point cloud dataset $\mathcal{D}$ consisting of $n$ points with coordinates $\mathcal{D}=\{\bm{x}_1,\bm{x}_2,\ldots,\bm{x}_n\}$ specified in Euclidean space $\mathbb{R}^d$. We aim to learn its topological properties~\cite{Lloyd_2016, Hayakawa_2022, Scali_2024, Scali_2024_b}, which can be difficult to access due to the combinatorial complexity for this task~\cite{Edelsbrunner_2009, Chazal_2021}. Inspired by the manifold reconstruction problem~\cite{Ho_Le_1988, Boissonnat_2013} and the simplicial finite element meshes in computational geometry~\cite{Mota_2007, Lee_2019}, we start by generating a regular (equilateral) triangle mesh of the $d$-dimensional Euclidean space on which the dataset is defined, i.e. a discrete local approximation of the continuous space $\mathbb{R}^d$. We construct a triangular mesh with edges of length equal to the average distance between data points. This choice, defining a trade-off between information loss due to space discretization and computational resources needed to run simulations, falls within the broader framework of mesh refinement~\cite{bank1983some}. We then map points of $\mathcal{D}$ to closest vertices of the mesh, generating an approximated version of the dataset referred to as $\tilde{\mathcal{D}}$ with new coordinates $\tilde{\mathcal{D}} = \{\tilde{\bm x}_1, \tilde{\bm x}_2, \ldots, \tilde{\bm x}_n \}$~\footnote{In this step, several optimizations and exceptions might arise~\cite{Bern_2000}, for example, the best mapping to minimize the overall approximation, multiple data points ending up on the same vertex in the mesh, etc. Since the problem is tangential to the main message of the paper, we assume that such unambiguous mapping exists and that it is a ``good-enough approximation'' for our needs.}. The approximated dataset can be associated with \v{C}ech simplicial complex $\Gamma$ \cite{wasserman2016tda} whose set of vertices (0-simplices) corresponds to the approximated coordinates and whose edges exist between vertices that are nearest neighbors on the underlying mesh. The complex is built relative to a filtration distance $\epsilon$ where the equilateral triangles in the mesh have edges of length $2\epsilon$~\cite{Chazal_2021}. Thus, the study of the topological properties of the original dataset $\mathcal{D}$ translates into the evaluation of $k$-th Betti numbers $\beta_k$ (i.e. the ``$k$-th order topological holes'') in the approximated dataset $\tilde{\mathcal{D}}$. 
In the following, we restrict the point cloud dimensionality to $d=2$, matching the pump profile in our simulations. The resulting mesh construction is a triangular lattice whose constant is the average spacing between data points of $\mathcal{D}$. 

First, we consider a problem of extracting the first-order Betti number $\beta_1$ of the simplicial complex associated with the approximated dataset, that is, its 2-dimensional holes. Note that in the regular triangle mesh described before, the evaluation of $\beta_1$ equates to solving the clique counting problem~\cite{Moon_1965}, where cliques are defined as subsets of vertices that are fully connected. We note that clique counting is NP-complete \cite{Karp1972} and contains instances that cannot be readily analyzed. In our case, this is framed as a learning problem for classifying point clouds with two and three cliques, as compared to samples with no cliques present.

Next, going beyond Betti number classification, we introduce an asymmetry detection task to evaluate the ability to capture subtle structural differences. We create binary datasets where both classes contain triangular patterns in a kagome lattice configuration (see SM, Fig.~\ref{fig_S1}). 
The key distinction between the two classes is a geometric transformation: one class maintains regular triangular spot configurations, while the second class features \textit{twisted} triangular spots. This twist is implemented by extending one spot by $2.0\,\mu$m along its centroid line, creating a subtle structural modification that breaks the symmetry required for vortex formation (as illustrated in SM, Fig.~\ref{fig_S2}). This task is particularly challenging since the geometric difference between classes is minimal and requires learning fine-grained geometric features with nearly degenerate configurations.
\begin{figure}[t]
\includegraphics[scale=1.0]{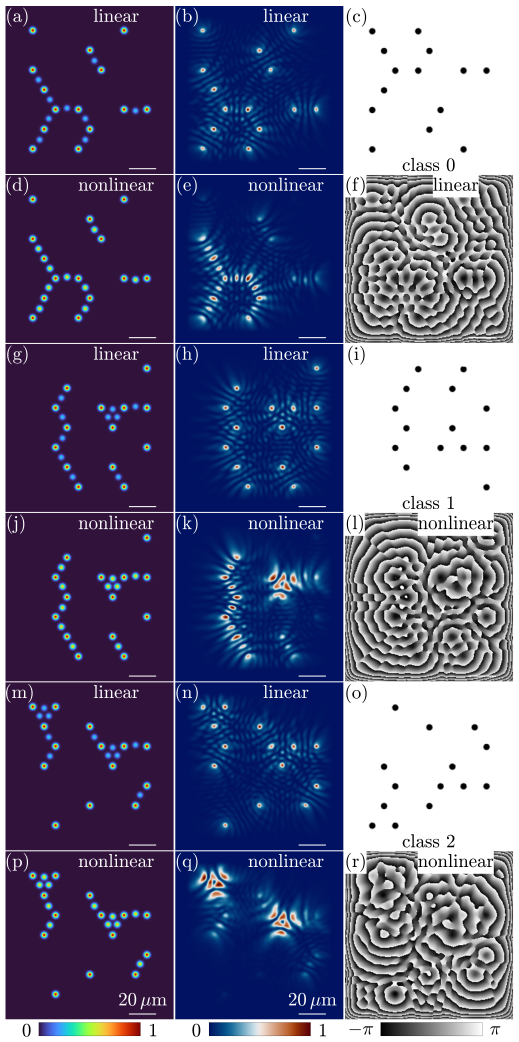}
\caption{Examples of pump profiles for linear (a,g,m) and nonlinear (d,j,p) photonic embedding of point clouds shown in (c,i,o). Photoluminescenece profiles are shown in (b,e,h,k,n,q), and the corresponding phase profile are presented in (f,l,r) for nonlinear operation. Topological features of polaritonic flow can be clearly observed in (k) and (q).}
\label{fig_app}
\end{figure}

\textit{Model: CNNs for geometric data.---}Convolutional neural networks, originally designed for grid-structured data such as images, can be adapted to process point clouds by stacking convolutional layers with learnable filters~\cite{lecun1989backpropagation, krizhevsky2012imagenet, LeCun2015}. These filters capture spatial hierarchies of features, enabling CNNs to recognize fundamental patterns such as edges, shapes, and textures. We use a CNN-based model with two convolutional layers (detailed in SM), each followed by batch normalization~\cite{ioffe2015batch} and max pooling operations. The depth is chosen such that for the limited sample size ($|\mathcal{D}| \leq 1280$ for two classes) CNNs do not overfit or suffer from training difficulties. The network processes $256 \times 256$ grayscale images using $3 \times 3$ kernels, ReLU activations, and global average pooling~\cite{lin2013network}, resulting in $83,331$ trainable parameters.

The full ML-based workflow for classification is shown in Fig.~\ref{fig_1}, visualizing analysis with different inputs.
\begin{table*}[t!]
\caption{\label{tab:table1}Mean and standard deviation of test accuracy for several classification tasks and different data processing.}
\begin{ruledtabular}
\begin{tabular}{c c c c c c c}
\textrm{\# classes (task)} & 
\textrm{\# samples} & 
\textrm{\# parameters}& \textrm{PL (polaritonics)} & \textrm{PL (photonics, $L=1$)} & 
\textrm{PL (photonics, $L=365$)} & 
\textrm{Graph images} 
\\
\colrule
$2\,\,(\text{clique det.})$ &$640$& $83,331$ & $\bm{0.8776 \pm 0.1008}$ & $ 0.8057 \pm 0.1304$ & 
$0.7500 \pm 0.0718$ & $ 0.5151\pm 0.0281$ 
\\
$3\,\,(\text{clique det.})$ &$640$& $83,331$ & $ 0.8931\pm 0.0167$ & $\bm{ 0.9330\pm0.0255 }$ & 
$ 0.9097\pm 0.0359$& $  0.3507\pm 0.0211$ 
\\
$2\,\,(\text{asym. det.})$ &$500$& $83,331$ & $ 0.8313\pm 0.0288$ & $ 0.7053\pm 0.0971 $ & 
$ \bm{0.9247\pm 0.0888}$& $  0.6893\pm 0.0411$ 
\\
\end{tabular}
\end{ruledtabular}
\end{table*}

\textit{Model: polaritonic lattices.---}We proceed to model nonequilibrium polariton condensates, forming a lattice-type structure. Once propagated in time, resulting photoluminescence patterns can be used similarly to positional embeddings in the ML workflow. The modeling can be performed by solving a generalized Gross-Pitaevskii equation (GPE), coupled to the exciton reservoir rate equation~\cite{carusotto2013quantum}. The nonlinear GPE reads~\cite{wouters2007excitations}
\begin{eqnarray}
i\hbar\frac{\partial }{\partial t}\Psi&=&\bigg\{-\frac{\hbar^{2}}{2m}\nabla^{2}
 + \alpha |\Psi|^2 
 + G\Big[\mathcal{N} 
+ \frac{\eta }{\Gamma}P(\bm{r})\Big]  \nonumber
\\ &&
+ i\frac{ \hbar}{2}\big[R \mathcal{N}-\gamma\big]\bigg\}\Psi,
\label{main_GPE}
\\
\frac{\partial }{\partial t}\mathcal{N}&=&-\Big[\Gamma+R |\Psi|^{2}\Big]\mathcal{N} + P(\bm{r}),
\label{rate_equation_reservoir}
\end{eqnarray}
where $m$ is an effective polariton mass (lower branch), $\alpha$ and $G$ denote the interaction strengths of polariton-polariton and polariton-reservoir, respectively. For the reservoir described by the number of excitons $\mathcal{N}(\bm{r},t)$, pumped continuously at rate $P(\bm{r})$, we introduce a scattering rate $R$ coupling reservoir and condensates. Here, $\eta$ is the ratio of dark excitons, and $\gamma$ ($\Gamma$) is decay rate of polaritons (reservoir). While various polaritonic systems can be used, we consider a planar stack of InGaAs quantum wells with negative detuning \cite{Alyatkin2021}.


The dataset is embedded via pump configurations specified as Gaussian spots following point clouds, $P(\bm{r}) = p\int \mathcal{L}(\bm{r})G(\bm{r}^{\prime})d\bm{r^{\prime}}$, where $\mathcal{L}(\bm{r}) = \sum_{i}\delta(\bm{r}-\bm{r}_{i})$ represents a sum of Dirac delta functions corresponding to selected points, and $p$ is a scalar pump rate. Each spot follows a Gaussian distribution, $G(\bm{r}) = (2\pi \sigma^{2})^{-1}\exp{[-|\bm{r}-\bm{r}^{\prime}|/(2\sigma^2)}]$. For those data spots nonresonantly excited above threshold, the coupling between spots introduces nonlinear features that depend on the cloud geometry. We note that possible solutions include topologically nontrivial excitations like quantized vortices~\cite{Lagoudakis2008} and solitons. Here, the vortices correspond to integer phase winding numbers emerging from quantum flow circulation~\cite{carusotto2013quantum} and highlighting relevant topological information. 
In addition to the source terms, we add sub-threshold pumps that can introduce barrier spots. These barriers represent a fixed part of embedding, appearing in the middle of the triangle edges, facilitating vortex creation while contributing to interference patterns. All details for the data embedding procedure and parameters are detailed in SM. 

Once the generalized GPE is solved (propagated until time $t_{\rm{f}} \gg 1/\gamma$), the resulting PL profile $|\Psi(\bm{r},t_{\rm{f})}|$ is used as an input for CNN. Other features include phase profile, $\rm{arg}(\Psi(\bm{r},t_{\rm{f}}))$. In Fig.~\ref{fig_app}, we show simulations for different pump profiles (and point clouds) with the corresponding PL and phase profiles.

\textit{Model: linear photonics.---}Together with nonlinear (polaritonic) feature embedding, we provide a recipe for purely photonic feature engineering, also serving as a reference point. Here, relevant features also appear within interference patterns and can yield the learning procedure (see Fig.~\ref{fig_1}). In this case, the underlying GPE reads
\begin{eqnarray}
&&i\hbar\frac{\partial }{\partial t}\Psi=\bigg[-\frac{\hbar^{2}}{2m}\nabla^{2}
 + V(\bm{r})-i\frac{\hbar}{2}\gamma\bigg]\psi  \nonumber
\\ &&
+ \frac{1}{L}\sum_{n=0}^{N-1}\sum_{l=0}^{L-1}E(\bm{r},\bm{r}_{0,n})e^{-i\big[\omega t-\bm{k}_{l}\cdot(\bm{r}-\bm{r}_{0,n})\big]},
\label{linear_photonic_equation}
\end{eqnarray}
where $V$ is a real potential of barrier spots, $N$ is the number of activated source spots, and other parameters remain the same as in Eq.~\eqref{main_GPE}). Since we do not have condensates in the linear case (exciton-exciton interaction is absent), the data embedding is modified and uses a set of resonant pumps. Specifically, $E$ is the profile map shaped into different normalized Gaussian spots at position $\bm{r}_{0,n}$ with the same FWHM as the nonresonant pump profile in the polaritonic case; $\omega$ represents the energy of the driving source; $\bm{k}_{l}=(2m\omega/\hbar)^{1/2}[\cos{(2\pi l/L)}\hat{\bm{x}}+\sin{(2\pi l/L)}\hat{\bm{y}}]$ is the resonant wave vector pointing at different directions defined by angles $2\pi/L$ for an integer $L$. Here, $(\hat{\bm{x}},\hat{\bm{y}})$ refers to the two-dimensional unit of $\bm{r}$. We consider two possible driving configurations. The first corresponds to a multidirectional resonant driving ($L=365$), closer to condensate-based processing from the feature engineering perspective, while more challenging experimentally (many driving lasers are needed). The second configuration corresponds to unidirectional driving ($L=1$) that is easier to implement in practice, but requires breaking the concentric symmetry. Simulation parameters for the linear photonic model are specified in SM.


\textit{Results.---}We apply the proposed polariton-enriched CNNs for classification tasks based on clique detection and asymmetry detection. For clique detection, we test both binary ($2$-class) and $3$-class classification settings, with configurations containing zero to one triangle and zero to two triangles, respectively. For the clique detection and asymmetry detection tasks, we use $640$ and $500$ samples per class, respectively. In each case, the results were averaged over $10$ independent runs using different random seeds. The dataset is split into training, validation, and testing subsets with ratios of $70\%$, $15\%$, and $15\%$, respectively. Stratified sampling is used to maintain class balance. Our results are summarized in Table~\ref{tab:table1}.

For the $2$-class clique detection task, polaritonic PL provided the highest test accuracy of $87.76\% \pm 10.08\%$, substantially outperforming linear photonic systems ($L=1$: $80.57\% \pm 13.04\%$; $L=365$: $75.00\% \pm 7.18\%$) and pure graph images ($51.51\% \pm 2.81\%$). This boost in performance can be traced to the formation of vortices in polaritonic systems, where topological defects appear when pump spots form a clique-shape configuration (e.g. triangle). These vortices, generated by the nonlinear interaction term in the generalized GPE, serve as a signature of clique presence. The coherent superfluid flow between condensate spots further creates distinct wake patterns that encode graph connectivity information into time-integrated PL images. In the $3$-class clique detection task, the performance hierarchy shifts: the linear photonic system with $L=1$ achieves the best accuracy at $93.30\% \pm 2.55\%$, followed by the case with $L=365$ ($90.97\% \pm 3.59\%$) and the polaritonic system ($89.31\% \pm 1.67\%$). This reversal indicates that when distinguishing between multiple clique configurations, the directional interference patterns generated by $L=1$ pump configuration provide features that are easier to digest. The asymmetric intensity distribution effectively encodes the number and arrangement of triangles, while the isotropic $L=365$ case generates concentric interference. All photonic configurations largely outperform the baseline of point cloud images ($35.07\% \pm 2.11$).

For the binary asymmetry detection task, which requires distinguishing between regular and twisted triangular configurations (with one spot extended by $2.0\,\,\mu\mathrm{m}$), the photonic system with $L=365$ achieves the highest accuracy at $92.47\% \pm 8.88\%$. The polaritonic system follows with $83.13\% \pm 2.88\%$, while the photonic system with $L=1$ ($70.53\% \pm 9.71\%$) performs only marginally better than graph images ($68.93\% \pm 4.11\%$). The sensitivity of vortex formation to geometric symmetry explains the slightly lower performance of polaritonic embedding, where the twisted configuration disrupts the vortex formation in the kagome geometry. Isotropic interference patterns of the $L=365$ system are most effective at capturing subtle geometric asymmetries.

\textit{Discussion.---}Results of the photonics-based ML demonstrate that incorporating physical features significantly enhances CNN performance for graph-based tasks compared to using raw graph representations. The evolution of nonlinear photonic systems and their photoluminescence profiles can be seen as an effective feature engineering for graphs embedded as pump profiles. In polaritonic systems with nontrivial many-body dynamics, the generation of quantized vortices emerges as the key nonlinear feature, highlighting cliques and topological structures. At the same time, we note that features arising from interference (also present in linear photonic systems) can be suitable for some classification tasks.

We note that both 2-class clique detection and asymmetry tasks are characterized by a large standard deviation (e.g. $\pm 10.08\%$ for polaritonics). This is due to the limited number of training samples ($640$ per class) combined with high sample diversity. However, polaritonic systems show remarkably low deviation for the $3$-class task ($\pm 1.67\%$), indicating that nonlinear features are reliable when the task complexity increases. For the asymmetry detection, the $L=365$ photonic system outperforms polaritonics ($92.47\%$ versus $83.13\%$), though with a larger standard deviation ($\pm 8.88\%$ versus $\pm 2.88\%$). Here, the twisted configuration disrupts vortex formation, while isotropic interference patterns of $L=365$ capture this geometric asymmetry more effectively. The poor performance of $L=1$ photonics ($70.53\%$) for this task, comparable to graph images ($68.93\%$), indicates that unidirectional flow provides insufficient information for detecting subtle geometric distortions.

The computational implications of our findings are significant: by mapping the NP-complete clique counting problem onto photonic systems, we demonstrate how physics-based positional embedding can tackle combinatorially hard problems with limited data. The picosecond-scale dynamics of polariton condensation further suggests potential for ultrafast graph processing applications, as well as reduced energy consumption \cite{Matuszewski2021}.

\textit{Conclusions.---}We demonstrated that polaritonic systems can be used for machine learning by performing photonic feature engineering based on their nonlinear and topological properties. We showed that lattices of condensates can encode relational and topological information from point cloud data in a physically meaningful way. When these photonic features were used as input to convolutional neural networks, the resulting models achieved around 90\% accuracy in Betti number classification and clique detection tasks, compared to 35\% accuracy from models without photonic preprocessing. These results show that nonlinear photonic systems can act as efficient and task-relevant feature processors for machine learning pipelines, enabling faster and more accurate processing of graph-structured data.

\begin{acknowledgments}
\textit{Acknowledgments.---}The authors acknowledge the support from UK EPSRC grant EP/X017222/1.
\end{acknowledgments}

\textit{Code availability.---}Code and results will be available on GitHub after journal submission.

%

\clearpage

\newpage

\appendix

\setcounter{figure}{0}
\renewcommand{\figurename}{Fig.}
\renewcommand{\thefigure}{S\arabic{figure}}

\section{SUPPLEMENTAL MATERIAL}

\section{Datasets preparation}
\label{appendix:datasets}
\begin{figure}[b]
\includegraphics[scale=1.0]{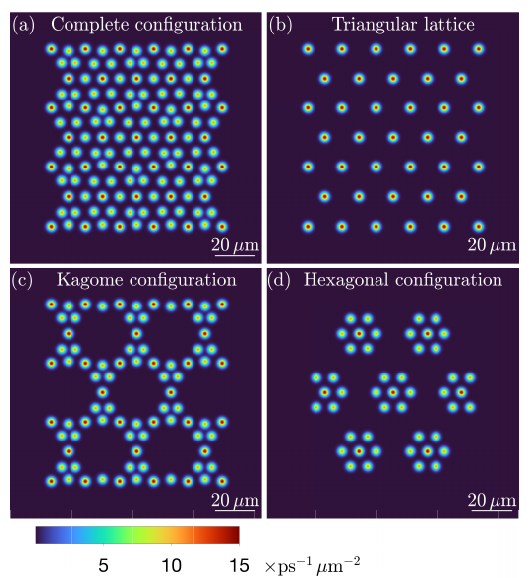}
\caption{(a) Complete pump configuration is produced with (b) triangular lattice made of only source spots and the rest barrier spots. This configuration can also be formed by combining a (c) kagome pattern with additional (d) hexagonal configuration.}
\label{fig_S1}
\end{figure}  

We consider a planar stack of InGaAs quantum wells with slightly negatively detuned cavities \cite{Alyatkin2021}. The detuning between the exciton and photon modes controls the interaction terms as $\alpha=g|\chi|^4$ and $G=2g|\chi|^2$, with $g=g_0/N$, where $g_0$ is the exciton-exciton interaction strength, and $N_{\text{QW}}$ is the number of quantum wells (QWs). The term $|\chi|^2$, representing the fraction of exciton in the polariton, is defined by the Hopfield coefficient~\cite{hopfield1958theory} for the excitonic branch.

The parameters chosen to emulate the experimental observations of polariton condensates are $m=0.28\mathrm{\,meV\,ps^{2}\,\mu m^{-2}}$, $|\chi|^{2}=0.4$, $N_{\text{QW}}=6$, $g_{0}= 0.01 \,\mathrm{meV\,\mu m^{2}}$, $\hbar R=10g$, $\eta=2$, and $\gamma^{-1} = \Gamma^{-1} = \SI{5.5}{ps}$. The power density $p$ of the source and barrier for all simulations of polariton condensation is $15\,\,\mathrm{ps}^{-1}\mathrm{\mu m}^{-2}$ and $11\,\,\mathrm{ps}^{-1}\mathrm{\mu m}^{-2}$, respectively. For simulation of the linear photonic system, $\hbar\omega=0.6\,\,\mathrm{meV}$, and the highest intensity of the real-valued Gaussian-distributed potential $V(\bm{r})$ is fixed at $0.3\,\,\mathrm{meV}$. 
\begin{figure}[b]
\includegraphics[scale=1.0]{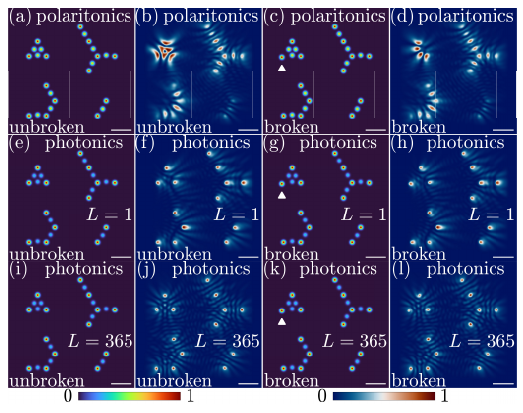}
\caption{Comparison of PL images for (a,e,i) unbroken and (c,g,k) broken pumping configurations in the (b,d) nonlinear polaritonic system, (f,h) linear photonic system with $L=1$, and (j,l) linear photonic system with $L=365$. White triangle marks the twisted spot in the broken configuration. The spot is extended by $2.0\,\,\mathrm{\mu m}$ along its centroid line, resulting in a twisted configuration. White scale bar marks $20.0\,\,\mathrm{\mu m}$.}
\label{fig_S2}
\end{figure}

For pump configuration, $\sigma \approx 1.274\,\,\mathrm{\mu m}$, which corresponds to the full width at half maximum (FWHM) of each Gaussian spot being $3.0\,\,\mathrm{\mu m}$ is used for all the Gaussian spots in this paper. As shown in Fig.~\ref{fig_S1}, the geometric parameters are as follows.  For the kagome configuration, the center-to-edge spacing within the equilateral triangles is $10\,\,\mathrm{\mu m}$ for the source spots and $4\,\,\mathrm{\mu m}$ for the inner barrier spots. For the hexagonal units, the distance between a source spot and its nearest barrier spots is half of the triangular lattice constant, with the lattice constant itself being approximately $17.3\,\,\mathrm{\mu m}$. All datasets are presented as single-channel, grayscale images, and added to visualize relevant configurations.

The PL images for nonlinear polaritonic system and the phase images are obtained, respectively, from the time-integrated $|\Psi|^2$ and the phase of the final-time $\Psi$ through numerical solutions of the generalized GPE coupled with reservoir rate equation [see Eqs.~\eqref{main_GPE} and~\eqref{rate_equation_reservoir} of the main content]. PL maps for linear photonic system are obtained from the final-time solution $|\Psi|^2$ of Eq.~\eqref{linear_photonic_equation} from the main text. The simulations are performed using internally prepared GPU-executed for propagating generalized GPEs.

The machine learning results presented in Table~\ref{tab:table1} are based on datasets randomly sampled from $14$ of the $39$ available source points within the lattice. Each dataset per category comprises $640\times3 = 1920$ images with dimensions of $256 \times 256$ pixels per image, where $640$ is the total number of images per category per class and $3$ represents the total number of classes. All images used for training are in single-channel and grayscale format. These images include cloud points, phase maps, linear photonic PL, and nonlinear polaritonic time-integrated PL.


\section{Details of convolutional neural network  architecture}

We use a convolutional neural network with two layers (detailed in the code listing), each followed by batch normalization~\cite{ioffe2015batch} and max pooling operations. The architecture progressively increases feature maps from $64$ to $128$ channels while maintaining spatial resolution through padding. Each convolutional layer employs $3\times 3$ kernels with stride $1$ and padding $1$, followed by ReLU activation functions. The network processes $256 \times 256$ grayscale images, with feature map dimensions reducing from $128 \times 128$ after the first pooling layer to $64 \times 64$ after the second. To address the parameter efficiency required for small datasets, we incorporated global average pooling~\cite{lin2013network} before the fully connected layers, reducing the total trainable parameters to $83,331$. The model was trained using the Adam optimizer~\cite{kingma2014adam} with an adaptive learning rate scheduler (\textsf{ReduceLROnPlateau} with \textsf{factor}=0.8, \textsf{patience}=20).
\begin{lstlisting}[language=Python]
class CNN(nn.Module):
    def __init__(self, num_classes=2, dropout_rate=0):
        super(CNN, self).__init__()
        
        self.conv1 = nn.Conv2d(1, 64, kernel_size=3, stride=1, padding=1)
        self.bn1 = nn.BatchNorm2d(64)
        
        self.conv2 = nn.Conv2d(64, 128, kernel_size=3, stride=1, padding=1)
        self.bn2 = nn.BatchNorm2d(128)
        self.max_pool = nn.MaxPool2d(kernel_size=2, stride=2)
        self.relu = nn.ReLU()
        self.dropout_conv = nn.Dropout2d(p=0)  
        self.dropout_fc = nn.Dropout(p=dropout_rate)
        self.global_avg_pool = nn.AdaptiveAvgPool2d(1)
        
        self.fc1 = nn.Linear(128, 64)
        self.fc2 = nn.Linear(64, num_classes)

    def forward(self, x):
        # Convolutional block 1
        x = self.conv1(x)
        x = self.bn1(x)
        x = self.relu(x)
        x = self.max_pool(x)
        x = self.dropout_conv(x)  
        
        # Convolutional block 2
        x = self.conv2(x)
        x = self.bn2(x)
        x = self.relu(x)
        x = self.max_pool(x)
        x = self.dropout_conv(x)  
        
        x = self.global_avg_pool(x)
        x = x.view(x.size(0), -1)
        
        x = self.relu(self.fc1(x))
        x = self.dropout_fc(x)  
        x = self.fc2(x)
        return x
\end{lstlisting}
Listing above shows the exact code of the CNN architecture in PyTorch.

\end{document}